\documentclass[superscriptaddress]{revtex4}
\usepackage[dvips]{graphicx}
\usepackage{bm}

\begin{document}

\title{Sealing is at the Origin of Rubber Slipping on Wet Roads} 
\author{B.N.J. Persson}
\affiliation{IFF, FZ-J\"ulich, 52425 J\"ulich, Germany}
\author{U. Tartaglino}
\affiliation{IFF, FZ-J\"ulich, 52425 J\"ulich, Germany}
\affiliation{International School for Advanced Studies (SISSA),
Via Beirut 2, I-34014 Trieste, Italy}
\affiliation{INFM Democritos National Simulation Center,
Via Beirut 2, I-34014, Trieste, Italy}
\author{O. Albohr}
\affiliation{Pirelli Deutschland AG, 64733 H\"ochst/Odenwald, Postfach 1120, Germany}
\author{E. Tosatti}
\affiliation{International School for Advanced Studies (SISSA),
Via Beirut 2, I-34014 Trieste, Italy}
\affiliation{INFM Democritos National Simulation Center,
Via Beirut 2, I-34014, Trieste, Italy}
\affiliation{International Center for Theoretical Physics (ICTP),
P.O.Box 586, I-34014 Trieste, Italy}

\maketitle

{\bf
Loss of braking power and rubber skidding on a wet road is
still an open physics problem, since neither the hydrodynamical effects
nor the loss of surface adhesion that are sometimes blamed really manage to 
explain the $20-30\%$ observed loss of low speed tire-road friction.
Here we advance a novel mechanism based on sealing of water-filled
substrate pools by the rubber. The sealed-in water effectively smoothens 
the substrate, thus reducing the viscoelastic dissipation in 
bulk rubber induced by surface asperities, well established 
as a major friction contribution. Starting with the measured
spectrum of asperities one can calculate the water-smoothened spectrum
and from that the predicted friction reduction, which is of the right magnitude.
The theory is directly supported by fresh tire-asphalt friction data.
}
\vskip 0.3cm

First principle calculations of frictional forces for realistic systems 
are generally impossible. The reason is that friction 
usually is an interfacial property, often determined by the last few 
uncontrolled monolayers of atoms or molecules at the interface. An extreme 
illustration of this is offered by diamond. The friction between two clean 
diamond surfaces in ultra high vacuum is huge because of the strong 
interaction between the surface dangling bonds. However, when the
dangling bonds are saturated by a  hydrogen monolayer
(as they generally are in real life conditions), friction becomes 
extremely low\cite{Flipse}. Since most surfaces of practical use are 
covered by several monolayers of contaminant molecules of unknown 
composition, a quantitative prediction of sliding friction coefficients
is out of the question. An exception to this may be rubber friction 
on rough surfaces, which is the subject we address here.

Rubber friction is of extreme practical importance, e.g., in the context 
of tires, wiper blades, conveyor belts and sealings\cite{Moore}.
Rubber friction on smooth substrates, e.g., on smooth glass 
surfaces, has two contributions, namely an adhesive (surface) and a hysteretic 
(bulk) contribution\cite{Moore,PerssonSS}. The adhesive contribution
results from the attractive binding forces between the rubber 
surface and the substrate. Surface forces are often dominated by 
weak attractive van der Waals interactions.
For very smooth substrates, because of the low elastic moduli of rubber-like materials, 
even when the applied squeezing force is very gentle this weak attraction 
may result in a nearly complete contact at the interface\cite{PerssonEur,Oliver}, 
leading to the large sliding friction force usually observed\cite{Grosch}. 
For rough surfaces on the other hand the adhesive contribution to rubber 
friction will be much smaller because of the small contact area. The actual 
contact area between a tire and the road surface, for example, is 
typically only $\sim 1 \%$ of the nominal footprint contact area\cite{PerssonJCP,Heinrich}. 
Under these conditions the bulk (hysteretic) friction mechanism 
is believed to prevail. The substrate asperities exert pulsating forces 
on the rubber. A large internal friction at the appropriate frequencies 
causes a large energy dissipation in the rubber bulk, and that in turn
provides the main source of friction\cite{PerssonJCP,Heinrich}.
For example, the exquisite sensitivity of tire-road friction to temperature 
just reflects the strong temperature dependence of the viscoelastic bulk 
properties of rubber. 

The bulk nature of rubber-rough substrate friction
is also the key that permits its quantitative characterization and calculation.
In fact, we showed recently that the observed sliding friction of a tire 
on a dry road surface can be calculated quite accurately by assuming it to 
be due entirely to internal damping in the rubber\cite{PerssonJCP}.
This theory takes into account the pulsating forces acting on the rubber 
surface from road asperities from all length scales, from  
$\lambda_0 \sim 1 \ {\rm cm}$, corresponding to the largest road asperities, 
down to micro-asperities characterized by a wavelength
$\lambda_c$ of order $\sim 1\div 10 \ {\rm \mu m}$ (theory shows that shorter 
wavelength roughness is unimportant). It gives friction coefficients of 
order unity, as are indeed observed experimentally.

\begin{figure}[htb]
  \includegraphics[width=0.30\textwidth]{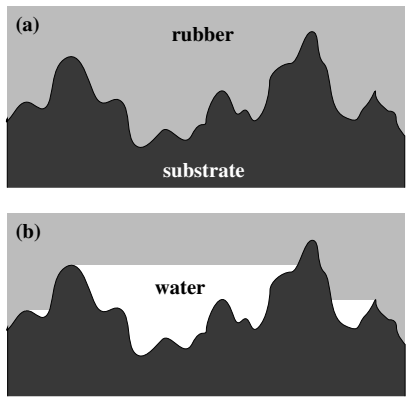}
  \caption{
A rubber block sliding on a rough hard substrate. (a) On a dry substrate 
the rubber penetrates  a large valley and explores the short wavelength 
roughness within. The pulsating rubber deformations induced by the short-wavelength 
roughness contributes to the friction force. 
(b) On a wet substrate the valley turns into a water pool. Sealing of the 
pool prevents the rubber from entering the valley. By removing
the valley contribution to the frictional force, the sealing effect
reduces the overall sliding friction.
    }
  \label{sealing}
\end{figure}

Our focus here is on rubber friction on {\em wet} rough substrates,
where at low sliding velocities it is known that the friction 
typically drops by as much as $20$--$30\%$ relative to the dry case\cite{Gert,MeyerWalter}. 
Owing to the small contact area, this cannot be the result of a
water-induced change of adhesion. On the other hand, as will be discussed below,
the friction decrease cannot be 
blamed on a purely hydrodynamical effect either.
That leaves finally
the possibility that water might change precisely the bulk, hysteretic friction. We propose 
here that this is indeed the case. Water pools that form in the wet 
rough substrate are {\em sealed off} by the rubber, as sketched in Fig.\,\ref{sealing}, 
and that will effectively smoothen the substrate surface. Smoothening reduces 
the viscoelastic deformation from the surface asperities, and thus reduces 
rubber friction.

\begin{figure}[htb]
  \includegraphics[width=0.35\textwidth]{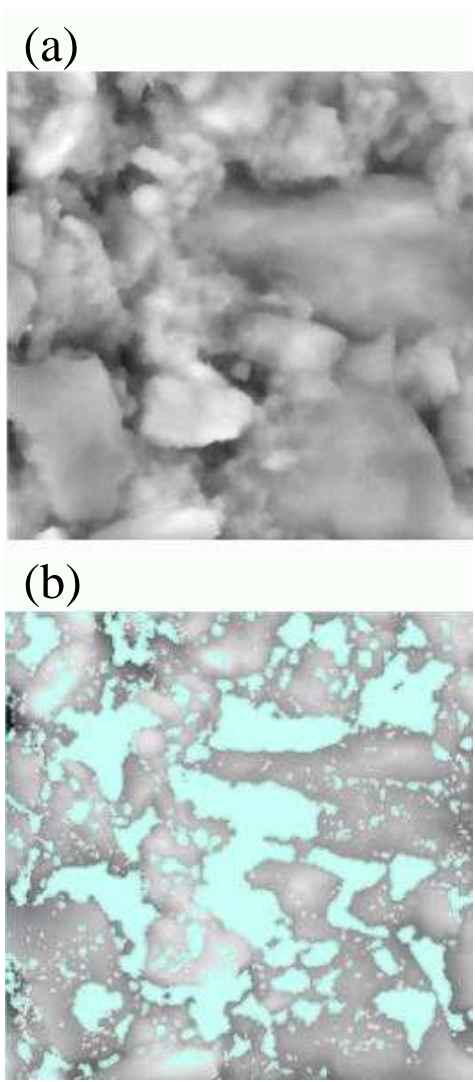}
  \caption{
Water smoothening of road roughness. (a) Optically measured height profile 
of a dry asphalt road ($1.5 \ {\rm cm} \times 1.5 \ {\rm cm}$ area). Darker 
areas correspond to deeper regions. (b) Calculated wet profile for the same 
area, with water pools (light blue).
}
  \label{topo}
\end{figure}

Rubber friction from the viscoelastic deformation by the substrate asperities 
is determined by the complex frequency-dependent bulk viscoelastic modulus 
$E(\omega )$ of rubber and by the substrate surface roughness power spectrum 
$C(q)$. If $h({\bf x})$ is the substrate height profile (measured from the average 
surface plane $\langle h \rangle = 0$), then
$$ C(q) = \frac{1}{(2\pi )^2}
\int d^2 x \ \langle h({\bf x}) h({\bf 0}) \rangle e^{-i{\bf q}
\cdot {\bf x}}. \eqno(1)$$
Fig.\,\ref{topo}(a) presents the height contour lines measured for a $1.5 \ {\rm cm}\times
1.5 \ {\rm cm}$ square of dry asphalt road. 

\begin{figure}[htb]
  \includegraphics[width=0.35\textwidth]{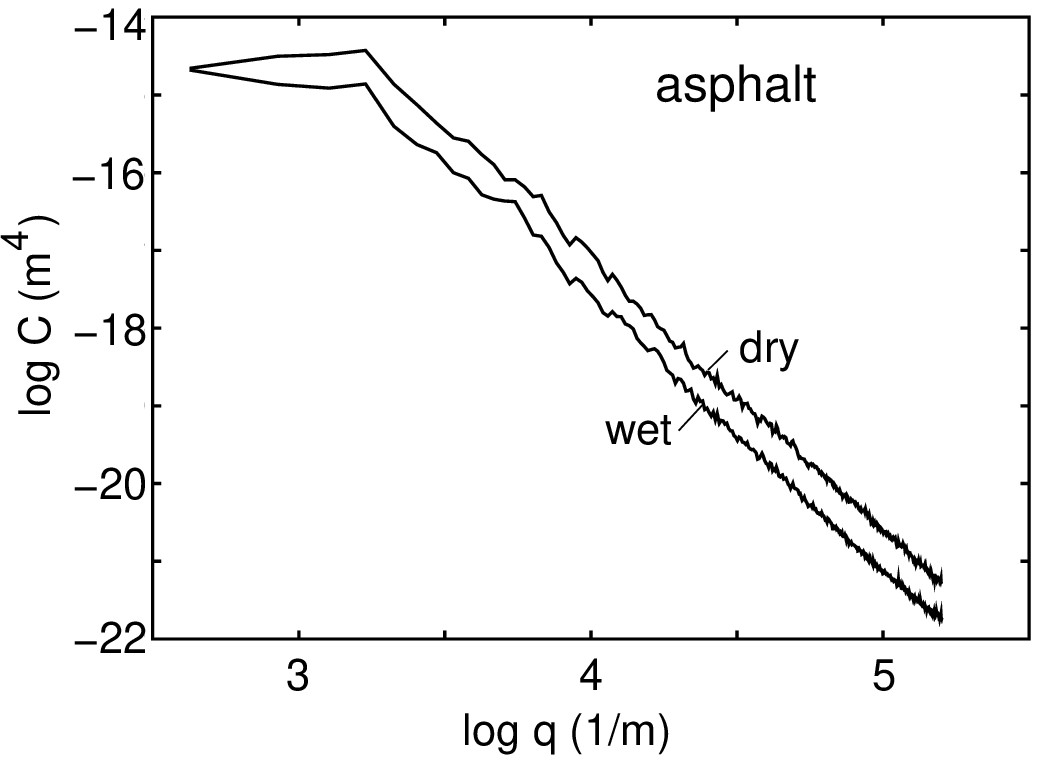}
  \caption{
Surface roughness power spectra $C(q)$ (above) extracted
from the measured height profile for a dry asphalt road surface, and (below) 
calculated assuming sealing of all pools in the same surface when wet, as in Fig.\,1.
Note the logarithmic scales.
    }
  \label{Cq}
\end{figure}

Assuming isotropic and translationally invariant statistical properties for the 
substrate, $C(q)$ will only depend on the magnitude $q=|{\bf q}|$ of the 
wavevector ${\bf q}$. The upper curve in Fig.\,\ref{Cq} shows the
power spectrum extracted via Eq. (1) from the measured height profile $h({\bf x})$. 
The log-log scale shows that for $q > 1600 \ {\rm m}^{-1}$, $C(q)$ drops
as a power law, as expected for a self-affine fractal surface. The fractal 
dimension of this surface is determined by the slope 
of the curve in Fig.\,\ref{Cq} (which equal $2D_{\rm f}-8$, where
$D_{\rm f}$ is the fractal dimension) and is about $D_{\rm f} =2.2$, 
and the root-mean-square roughness can be obtained directly from the height
profile, 
$h_{\rm rms} \approx 0.3 \ {\rm mm}$. 

The roll-off wavevector $q_0=1600~{\rm m}^{-1}$
below which $C(q)$ is constant corresponds to the size
$\lambda_0 = 2\pi /q_0 \approx 4 \ {\rm mm}$ of the
largest asperities or sand particles, etc., contained in the asphalt.    
Hysteretic rubber friction will of course decrease with decreasing magnitude 
of $C(q)$. Depending on the sliding velocity $v$ it will also sample
differently the wavevector dependence of $C(q)$ over a wide range of $q$. 
For example, rubber friction on asphalt surfaces at typical car velocities
depends on $C(q)$ for $q_1 < q < q_2$, where $q_1\approx 10^3 \ {\rm m^{-1}}$ 
and $q_2 \approx 10^6 \ {\rm m}^{-1}$. 

Consider now a tire rolling and sliding on a wet road surface.
At low velocities (say 
$v < 60 \ {\rm km/h}$) there will be negligible hydrodynamic water 
buildup between the tire and the road surface.
In essence\cite{tobepubl},
if $v < (\sigma / \rho )^{1/2}$, where $\sigma$ is the perpendicular 
stress in the tire-road contact area and $\rho$ the water mass density,  
there is sufficient time for the water to be squeezed out
of the contact regions between the tire and the road surface, 
except for water trapped in road cavities. The water pools 
will be sealed off by the road-rubber contact at the upper boundaries of the 
cavities (see Fig.\,\ref{sealing}). Thus, in what follows we will 
focus on the smoothening effect on the road profile caused by the sealing.

Starting from a dry substrate profile $h({\bf x})$ we can 
numerically build a new wet surface height profile $h'({\bf x})$ 
as shown in Fig.\,\ref{topo}(b). The algorithm assumes every valley 
to be filled with water up to the maximum level where the water will 
remain confined, i.e., up to the lowest point of the edge surrounding 
the pool. The square size $D$ is 
assumed to be that of a tread block, so that
any extra water added to the 
profile of Fig.\,\ref{topo}(b) will flow straight out of the square area.
If $ D > \lambda_0$, as is the case in our data, this construction is 
unique. (Even if $ D < \lambda_0$ uniqueness can still be attained 
at the price of averaging over many realizations).

 From the water-smoothened height profile $h'({\bf x})$ we obtain a
modified power spectrum $C'(q)$ shown by the lower curve in Fig.\,\ref{Cq}.
While the fractal power-law decay and the roll-off wavevector are 
essentially the same as for the dry surface, the reduction in the power spectrum 
reflect the effective water-induced smoothening 
of the rough substrate.

The sealed-off water in the pools (see Fig.\,\ref{sealing}) 
removes contact with the 
interior of the valley, which smoothens the effective substrate roughness 
profile. Our basic assumption is therefore that when rubber 
slides on the wet rough surface, the friction force will be determined 
by the modified power spectrum $C'(q) < C(q)$.

Let us now examine, based on this model, numerical results on tire 
friction on dry and wet substrates, calculated using the hysteretic friction theory presented 
in Ref. \cite{PerssonJCP}. 
The hysteretic friction coefficient at velocity $v$ is determined
by knowledge of the rubber viscoelastic modulus $E(\omega )$ and of the surface 
roughness spectrum $C(q)$ by
$$\mu_{\rm k} = \frac{1}{2}\int dq \ q^3 C(q) P(q)
\int_0^{2\pi} d\phi \ {\rm cos} \phi \ {\rm Im}
\frac{E(qv \ {\rm cos}\phi )}{(1-\nu^2)\sigma},$$
where \textrm{Im} denotes the imaginary part of the complex Young modulus $E$
and $\phi$ is the angle between ${\bf q}$ and ${\bf v}$. Moreover
$$P(q)= \frac{2}{\pi} \int_0^\infty dx \ \frac{\sin x}{x} 
{\rm exp}\left [-x^2G(q) \right ]= {\rm erf}\left(1/2\surd G\right ),$$
where \textrm{erf} is the error function, and
$$G(q)=\frac{1}{8}\int_0^q dq \ q^3C(q)\int_0^{2\pi}d\phi \ \left|\frac{E(qv \ {\rm cos}\phi )}{
(1-\nu^2)\sigma}\right|^2,$$
where $\sigma$ is the mean perpendicular pressure (load divided by the 
nominal contact area), and $\nu $ the Poisson ratio, 
which equals 0.5 for rubber-like materials. 

We obtained in this way quantitative results for the friction
of a standard tread compound, sliding on 
the asphalt road just characterized. We used the measured rubber complex 
viscoelastic modulus (not shown)
along with the power spectra presented in Fig.\,\ref{Cq} for the dry and for
the wet road surfaces.

\begin{figure}[htb]
  \includegraphics[width=0.35\textwidth]{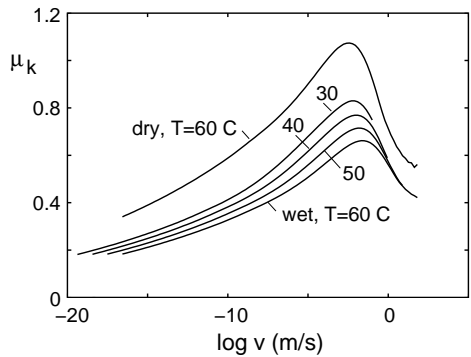}
  \caption{
Kinetic friction coefficient as a function of the logarithm of
the sliding velocity, calculated for a standard tread compound 
and an asphalt substrate with the roughness spectra of Fig.\,3.
    }
  \label{mukinetic}
\end{figure}
 
Fig.\,\ref{mukinetic} shows the rubber-asphalt kinetic friction coefficient
calculated for the dry surface 
at $T=60^\circ \ {\rm C}$ as a typical 
tire temperature while rolling on a dry road, and for the wet surface 
at four different temperatures, namely $T=30$, $40$, $50$, and
$60^\circ \ {\rm C}$ (on a wet road the tyre temperature is
typically $30^\circ \ {\rm C}$, and generally lower than on the dry surface).
Calculations include the flash temperature generated by the 
local energy dissipation,
shifting the rubber viscoelastic spectra 
toward the rubbery region. This effect
is important at sliding velocities $v > 1 \ {\rm cm/s}$.
Also, the decreasing friction with increasing 
temperature shown in Fig.\,\ref{mukinetic} is commonly observed for rubber.
The viscoelastic spectrum shifts to higher frequencies 
with increasing temperature, making rubber more elastic 
and less viscous, and reducing friction. 

When dry and wet
frictions are compared, the calculation shows first of all a 
water-induced friction decrease of $\sim  30\%$ at $T=60^\circ \ {\rm C}$. 
The decrease becomes somewhat less ($\sim  20\%$)
if the wet substrate temperature is (realistically) reduced to $T=30^\circ \ {\rm C}$.
We also calculated $\mu$-slip curves\cite{Gert,MeyerWalter},
which show a very similar reduction in the friction for wet road surfaces,
in excellent agreement with the known reduction
of low-speed rubber friction on road surfaces.

The above picture does in our view catch an important novel effect of water
on rubber friction. Yet, it is open to refinements in various 
ways. First, dry friction of tires is not pure sliding but also
involves some stick-slip\cite{lowvelocities}. This effect is included in
the calculation of $\mu$-slip curves, 
and the observed reduction in the effective friction
is similar to that in Fig. \ref{mukinetic}.  
Second, after enough time all sealings leak. This will be 
particularly true in the present case because the upper boundary of a 
water filled pool, which is in contact with the rubber, still has 
roughness on many length scales. So one cannot expect the rubber to 
make equally perfect contact everywhere, and there will be narrow channels 
through which the water slowly seeps out of the pools. As a result, for 
sufficiently low sliding velocities the negative water influence on rubber 
friction may revert to negligible. Experiments indeed show that to be the case 
for extremely low speed below $0.7 \ {\rm m/s}$, where the difference in 
$\mu_{\rm k}$ between dry and wet surfaces is very small\cite{lowvelocities}. 
We should also stress that the effects addressed here clearly apply only to moderately 
wet substrates and for speeds below $60 \ {\rm km/h}$.
For flooded surfaces, or for tires without a tread profile
and speeds above $60 \ {\rm km/h}$ aquaplaning may occur, 
which originates instead from the inertia of the water. Finally, for 
rubber friction on relative smooth wet surfaces, 
where the adhesional interaction is important, the so called dewetting
transition\cite{dewetting1,dewetting2,dewetting3} may be important.

Work in SISSA was sponsored by the Italian Ministry of University and
Research through MIUR COFIN 2003, and MIUR FIRB RBAU01LX5H as well as
through Istituto Nazionale Fisica della Materia, grant INFM FIRB
RBAU017S8R.
We very recently learnt that related unpublished ideas have also been
expressed by G. Heinrich, S. Kelbch, M. Klueppel and E. J. Schramm in Dresden.

\end{document}